\documentclass[prd,preprintnumbers,12pt]{revtex4}
\usepackage{epsfig}
\usepackage{graphicx}
\usepackage{bm}

\newcommand{\be}{\begin{equation}}

\newcommand{\ee}{\end{equation}}
\newcommand{\bea}{\begin{eqnarray}}
\newcommand{\eea}{\end{eqnarray}}

\def\qq{Q\!\!\!\!Q}

\newcommand{\nn}{\nonumber}
\newcommand{\de}{\partial}

\begin{document}

\preprint{{\bf BARI-TH 490/04}} \preprint{{\bf DFF-418/06/04}}
\title{Skyrmions and pentaquarks   in the quark-hadron continuity perspective}

\author{R. Casalbuoni}\email{casalbuoni@fi.infn.it}
\affiliation{Department of Physics, University of Florence, and
INFN-Florence , Italy}
\author{G. Nardulli}\email{giuseppe.nardulli@ba.infn.it}
\affiliation{Department of Physics, University of Bari and
INFN-Bari, Italy}

\begin{abstract} We argue that in
the color-flavor-locking (CFL) superconducting phase classical
soliton solutions can exist, whose excitations should be
interpreted as states  formed by a quark (or an antiquark) and
condensed diquarks. This finding extends the picture of
quark-hadron-continuity showing the existence of a region,
intermediate between the CFL and the hypernuclear phase, where
chiral solitons and Nambu Goldstone bosons can exist. We derive an
expression of the soliton mass in terms of the QCD coupling,
$g_s$, and  the Nambu Goldstone boson parameters. From the
quark-hadron continuity we can draw an argument in favor of the
interpretation  of the $\Theta^+(1540)$ particle in terms of a
strange antiquark and two highly correlated $ud$ pairs (diquarks).

\end{abstract}
 \maketitle
\section{Introduction \label{sec:0}}

The  observation of the baryon resonance $\Theta^+(1540)$ has been
recently reported by several groups. The results of the LEPS
\cite{Nakano:2003qx}, DIANA \cite{Barmin:2003vv}, CLAS
\cite{Stepanyan:2003qr,Kubarovsky:2003fi}, SAPHIR
\cite{Barth:2003es}, SVD \cite{Aleev:2004sa}, COSY-TOF
\cite{Abdel-Bary:2004ts}, ZEUS \cite{Chekanov:2004kn} and HERMES
\cite{Airapetian:2003ri} experiments as
 well as analyses of old bubble chamber experiments \cite{Asratyan:2003cb}
 show the existence of this narrow state ($\Gamma\sim$ a few MeV),
decaying into $K^+n$ or $K^0_sp$. The simplest quark model
interpretation is that of a pentaquark, i.e. an exotic state
formed by five quarks: $udud\bar s$. Other  narrow exotic cascade
states, e.g. a $\Xi^{--}$ state with quantum numbers $B=1,Q=S=-2$,
and also a $\Xi^{-}$ and $\Xi^{0}$ state have been reported by the
NA49 Collaboration, see \cite{Alt:2003vb}. Also these signals can
be interpreted as  pentaquark states, e.g. for $\Xi^{--}$,
$dsds\bar u$. Much experimental effort is expected in the near
future to consolidate these findings and clarify the experimental
problems. In any event the appearance of exotic states, coming
after years of fruitless experimental researches of exotica, has
revived theoretical interest in QCD spectroscopy and its low
energy models. Pentaquark states were indeed predicted long ago in
the framework of the chiral soliton model
\cite{Manohar:1984ys,Chemtob:1985ar}, which is an extension to
three flavors \cite{Witten:1983tx,Adkins:1983ya,Guadagnini:1984uv}
of the Skyrme model \cite{Skyrme:1961vq,Skyrme:1962vh}.  Its  mass
was also correctly predicted by \cite{BD} and
\cite{Praszalowicz:1987p}. In the chiral quark soliton model
\cite{Walliser:1992vx,Diakonov:1997mm} all baryonic states are
interpreted as arising from quantizing the chiral nucleon soliton
and the pentaquark emerges as the third rotational excitation with
states belonging to an antidecuplet with spin $s=1/2$. Other
interpretations have been proposed after the discovery of the
$\Theta^+(1540)$, most notably the one of Jaffe and Wilczek
\cite{Jaffe:2003sg,Jaffe:2004zg} who propose that the $\Theta^+$
comprises two highly correlated $ud$ pairs (diquarks: $\qq$) and
an $\bar s$. Diquark  properties are similar to those of the
diquark condensates of QCD in the high density
color-flavor-locking (CFL) phase \cite{Alford:1998mk}. The two
diquarks are in spin 0 state, antisymmetric in color and flavor.
Together they produce a $\qq\qq$ state in the flavor-symmetric
$\bf 6_f$ that must be antisymmetric in color and in $p-$wave to
satisfy Bose statistics. When combined with the antiquark the
diquarks produce a $\bf \overline{10}_f$ with spin 1/2 and
positive parity (they can also produce a $\bf 8_f$, and mixing is
possible).

The hypothesis that the attractive interaction in the
antisymmetric color channel may play a role both at low and high
density quark matter is especially interesting in the light of the
quark-hadron continuity which has been suggested
\cite{Schafer:1998ef} to exist between the CFL and the
hypernuclear phase.
 Due to the formation of the CFL condensate that breaks color,
 flavor and the electric charge, though conserving
a combination of the electric charge and of the color generator
$T_8$, the physical states are obtained by dressing the quarks by
diquarks. The result is that in this phase eight quarks have
exactly the same quantum numbers of baryons. Also the ninth quark
corresponds to a singlet with a gap which is twice the gap of the
octet. The same phenomenon takes place for the other states, as
for instance, the gluons in the CFL and the vector mesons in the
low density phase. In fact, the gluons are dressed by a pair
$\overline{\qq}\qq$ giving rise to vector states with the same
quantum number of the octet of vector resonances ($\rho$, etc.).
Also, the NG field $\phi$ associated with the breaking of $U(1)_V$
can be related to a possible light meson $H$ of the hypernuclear
phase \cite{Schafer:1998ef}. The state $H$ which is a six-quark
singlet of the type $udsuds$ was introduced by \cite{Jaffe:1977yi}
in the context of the bag model. A more detailed discussion of the
quark-hadron continuity can be found in \cite{Schafer:1998ef}.

Quark-hadron continuity plays a role in relating quark and baryons
in the low-lying octet. Apparently it also matters in assigning a
role to diquark attraction  at zero baryonic densities. In this
Letter we suggest that another sign of it is the possible
existence of baryon chiral solitons also at finite density. We
show below that they could arise in QCD at finite density by the
existence of a Skyrme term in the effective lagrangian for the
Nambu-Goldstone bosons of the CFL phase. The static solution of
the classical equations of motion has the  same form of the chiral
soliton model of refs.
 \cite{Adkins:1983ya,Guadagnini:1984uv} and
 \cite{Manohar:1984ys,Chemtob:1985ar}.
 Therefore its quantization will eventually produce baryonic
 states with properties similar to those of the low-lying baryonic
 octet as well as of its excitations, and in particular
 the pentaquark.

In Section II we discuss the effective lagrangian describing the
light modes of the CFL phase \cite{Casalbuoni:1999wu} and we show
that the decoupling of the gluons generate a Skyrme term. In
Section III we evaluate the soliton mass by extrapolating the
 parameters of the effective lagrangian down to chemical
 potentials of order $400\div 500~MeV$. We find a value of about
 $1200~MeV$ which is in the right ball-park. Also we evaluate the
 size of the soliton and we discuss the validity of our
 calculation. In Section
 III we discuss our results with a particular emphasis about the implications
 of the quark-hadron continuity idea on the pentaquark states.

\section{The effective lagrangian for the Goldstone bosons}
\label{sec:1}

We recall briefly the form of the effective lagrangian for the
light modes of the CFL phase. At this level the gluons should be
already decoupled since $p\ll\Delta$ and we know from
\cite{Casalbuoni:2000na} that the gluons in the CFL phase have
physical masses of order $\Delta$. However, since we want to show
that precisely the process of decoupling the gluon fields produces
the Skyrme term, we will write the effective action for the full
set of 18 Goldstone bosons from the breaking of $U(3)_L\otimes
U(3)_R\otimes SU(3)_c$ to $SU(3)\otimes Z_2\otimes Z_2$. The set
includes also the Goldstones to be eaten up by the gluon fields.
The effective lagrangian in this form has been discussed in
\cite{Casalbuoni:1999wu} (see also \cite{Hong:1999dk}). We can
associate the Goldstone fields to the left(right)-handed spin 0
diquark condensates according to   \be \hat X_\alpha^i\approx
\epsilon^{ijk}\epsilon_{\alpha\beta\gamma}\langle \psi^j_{\beta
L}\psi^k_{\gamma L}\rangle^*,~~~ \hat Y_\alpha^i\approx
\epsilon^{ijk}\epsilon_{\alpha\beta\gamma}\langle \psi^j_{\beta
R}\psi^k_{\gamma R}\rangle^*, \label{6.1}\ee with $\hat X$ and
$\hat Y$ $3\times 3$ unitary matrices. For the following it will
be more convenient to separate the $U(1)$ factors from $\hat X$
and $\hat Y$ by defining $U(1)$ and $SU(3)$ fields
 \be\hat X= X
e^{2i(\phi+\theta)},~~~\hat Y =Y e^{2i(\phi-\theta)},~~~ X, Y\in
SU(3). \label{6.4}\ee The transformation properties of these
fields under the full symmetry group are\be
 X\to g_c X g_L^T,~~~ Y\to g_c Y
g_R^T,~~~\phi\to\phi-\alpha,~~~\theta\to\theta-\beta.
\label{6.6}\ee with $\alpha$ and $\beta$ the parameters of the
groups $U(1)_V$ and $U(1)_A$ respectively and $g_c\in SU(3)_c$,
$g_{L,R}\in SU(3)_{L,R}$.

The breaking of the global symmetry can be discussed also using
the gauge invariant fields \be\Sigma^i_j=\sum_\alpha (
Y_\alpha^j)^*  X_\alpha^i\to \Sigma= Y^\dagger  X. \label{6.7}\ee
The $\Sigma$ field describes the 8 Goldstone bosons corresponding
to the breaking of the chiral symmetry $SU(3)_L\otimes SU(3)_R$,
as it is made clear by the transformation properties of
$\Sigma^T$, $\Sigma^T\to g_L \Sigma^T g_R^\dagger$. That is
$\Sigma^T$ transforms  as the usual chiral field. The other two
fields $\phi$ and $\theta$ provide the remaining Goldstone bosons
related to the breaking of the $U(1)$ factors. However, since the
$U(1)_A$ symmetry is anomalous although,  asymptotically in $\mu$,
gets restored, we will omit this field in the following
discussion.

 In order to build up an invariant lagrangian, it is
convenient to define the following currents \bea J_X^\mu&=&X D^\mu
X^\dagger= X(\partial^\mu X^\dagger+ X^\dagger
g^\mu)=X\partial^\mu X^\dagger+ g^\mu,\nn\\J_Y^\mu&=& Y D^\mu
Y^\dagger= Y(\partial^\mu Y^\dagger+ Y^\dagger g^\mu)=
Y\partial^\mu Y^\dagger+ g^\mu, \eea with \be g_\mu=ig_s g_\mu^a
T^a\ee the gluon field and \be T^a=\frac{\lambda_a}2\ee the
$SU(3)_c$ generators. These currents have simple transformation
properties under the full symmetry group $G$:\be J^\mu_{X,Y}\to
g_c J^\mu_{X,Y}g_c^\dagger.\ee The most general lagrangian, up to
two derivative terms, invariant under $G$, the rotation group
$O(3)$ (Lorentz invariance is broken by the chemical potential
term) and the parity transformation, defined as: \be P:~~~~
X\leftrightarrow Y,~~~ \phi\to\phi,\ee is \cite{Casalbuoni:1999wu}
\bea {\cal L}&=&-\frac{F_T^2}4{\rm
Tr}\left[\left(J^0_X-J^0_Y)^2\right)\right]-\alpha_T\frac{F_T^2}4{\rm
Tr}\left[\left(J^0_X+J^0_Y)^2\right)\right]+\frac 12
(\de_0\phi)^2+\nn\\
&&+\frac{F_S^2}4{\rm Tr}\left[\left|{\bf J}_X-{\bf
J}_Y\right|^2\right]+\alpha_S\frac{F_S^2}4{\rm Tr}\left[\left|{\bf
J}_X+{\bf
J}_Y\right|^2\right]-\frac{v_\phi^2}2|{\bm\nabla}\phi|^2- \frac
1{g_s^2}Tr[{\epsilon \bf E}^2-\frac 1{\lambda}{\bf
B}^2],\nn\label{lagrangian1}
 \eea or, in terms of the fields $X$ and $Y$
\bea {\cal L}&=&-\frac{F_T^2}4{\rm Tr}\left[\left( X\de_0
X^\dagger- Y\de_0
Y^\dagger)^2\right)\right]-\alpha_T\frac{F_T^2}4{\rm
Tr}\left[\left( X\de_0  X^\dagger+ Y\de_0
Y^\dagger+2 g_0)^2\right)\right]\nn\\
&&+\frac{F_S^2}4{\rm Tr}\left[\left|  X{\bm\nabla} X^\dagger-
Y{\bm\nabla} Y^\dagger\right|^2\right]+\alpha_S\frac{F_S^2}4{\rm
Tr}\left[\left| X{\bm\nabla} X^\dagger+ Y{\bm\nabla}\hat
Y^\dagger+2 {\bf g}\right|^2\right] \nn\\&&+\frac 12
(\de_0\phi)^2-\frac{v_\phi^2}2|{\bm\nabla}\phi|^2-\frac
1{g_s^2}Tr[{\epsilon \bf E}^2-\frac 1{\lambda}{\bf
B}^2],\label{lagrangian1}
 \eea where
 \be
 F_{\mu\nu}=\de_\mu g_\nu-\de_\nu g_\mu-[g_\mu,g_\nu],\ee
 and
 \be E_i=F_{0i},~~~~B_i=\frac 12\epsilon_{ijk}F_{jk}.\ee The parameters
 $\epsilon$ and $\lambda$ are the dielectric constant and the
 magnetic permeability of the dense condensed medium.

Notice that thee gluons $g_0^a$ and $g_i^a$ in the CFL vacuum
acquire Debye and Meissner masses given by \be m_D^2=\alpha_Tg_s^2
F_T^2,~~~m_M^2=\alpha_Sg_s^2 F_S^2=\alpha_Sg_s^2v^2 F_T^2,
\label{masses}\ee where we have introduced \be
v^2=\frac{F_S^2}{F_T^2}.\ee It should be stressed that these are
not the true rest masses of the gluons, since there is a large
wave function renormalization effect making the gluon masses of
the order of the gap $\Delta$, rather than $\mu$
\cite{Casalbuoni:2000na}. Since this description is supposed to be
valid at low energies  below the gap $\Delta$, we can decouple the
gluons solving their classical equations of motion neglecting the
kinetic term. The result from Eq. (\ref{lagrangian1}) is \be
g_\mu=-\frac 12 \left( X\de_\mu X^\dagger+ Y\de_\mu
Y^\dagger\right). \label{6.18}\ee By substituting this expression
in Eq. (\ref{lagrangian1}), and performing a gauge rotation to get
$Y=1$, one obtains \be {\cal L}=\frac{F_T^2}4\left({\rm
Tr}[\dot\Sigma\dot\Sigma^\dagger]-v^2{\rm
Tr}[\vec\nabla\Sigma\cdot\vec\nabla\Sigma^\dagger]\right)+ \frac
12\left(\dot\phi^2-v_\phi^2|\vec\nabla\phi|^2\right)-\frac
1{g_s^2}Tr[{\epsilon \bf E}^2-\frac 1{\lambda}{\bf B}^2],
\label{6.17}\ee where now \be E_i=\frac 1
4[\Sigma\de_0\Sigma^\dagger,\Sigma\de_i\Sigma^\dagger],~~~~~B_i=\frac
1
8\epsilon_{ijk}[\Sigma\de_j\Sigma^\dagger,\Sigma\de_k\Sigma^\dagger].\ee
Therefore, except for the breaking of the Lorentz symmetry, we
recognize in the first term the lowest order chiral lagrangian
and, in the last one,  the Skyrme term \cite{Skyrme:1962vh}.

It is interesting to notice that the idea of the Skyrme term
generated by decoupling the gauge boson of a hidden symmetry
\cite{Abud:1985wp} is realized here by decoupling the gluons (see
also \cite{Jackson:2003dk}). Notice also that one has to add to
this effective lagrangian the Wess-Zumino term as discussed in
\cite{Sannino:2000kg,Casalbuoni:2000jn}. It turns out that the
Wess-Zumino contribution coincides with the one at zero density
(see also \cite{Hsu:2000by}). The addition of this term is vital
for getting the right quantum numbers for the baryons once the
classical soliton solution is quantized
\cite{Witten:1983tx,Guadagnini:1984uv}.

\section{Numerical estimates\label{sec:2}}

Static solutions minimizing the energy are found assuming a
constant field $\phi$; for $\Sigma$ we make  the usual choice
\cite{Witten:1983tx} incorporating the hedgehog ansatz for the
SU(2) chiral subgroup \be \Sigma({\bf x})=\left(\matrix{ \exp[i(
{\bf x\cdot}{\bm \tau}F(r)/r  ]&0      \cr 0 &1 } \right), \ee
with $r=|{\bf x}|$, $F(0)=\pi$, $F(r)\to 0 $ when $r\to\infty$.
The soliton mass is a functional of $F(r)$ subject to the boundary
conditions given above. Minimization of the energy gives as a
result the usual relation between the parameters of the lagrangian
and the soliton mass \cite{Skyrme:1962vh}, \cite{Adkins:1983ya}.
In our case one should take into account a different normalization
of the pion decay constant and the pion velocity $v$, producing
$F_\pi\to 2F_T v$. Besides, the chromo-magnetic permeability
changes the coupling $g_s$ to $g_s\sqrt{\lambda}$. As a result we
get \be M=36.5\frac{2F_Tv}{g_s\sqrt \lambda}. \ee It should be
stressed that the soliton mass is given here in terms of the
parameters of the low-energy theory, $F_T$, $v$, the magnetic
permeability of the dense medium $\lambda$ and in terms of the
strong coupling constant $g_s$.  Therefore, at least in principle,
there are no free parameters and everything could be determined by
the fundamental theory. In fact, if we use the results of the
calculations at high density (see e.g.
\cite{Son:1999cm,Casalbuoni:2000na}) we get
  \be F_T^2=\frac{\mu^2(21-8\ln
2)}{36\pi^2}~,~~~v=\frac 1{\sqrt 3},\ee and
\bea\frac{1}{\lambda}&=&1+\frac{\mu^2g_s^2}{30\pi\Delta^2}(a+b)\ ,\nn\\
a+b&=& \frac{1}{108\pi}\left(41-\frac {112} 3\ln 2 \right). \eea
The gap is also determined by QCD at high density
\cite{Son:1998uk}.
 We can now
extrapolate this high-density prediction to values of $\mu$ of
order $400\div 600~MeV$ to get an idea of the order of magnitude
in a region that should be not too far from the hypernuclear
phase. Also we consider  $\Delta=40$ and $80~MeV$. For $g_s$ we
take $\sim 3.5$ corresponding to $g^2_s/4\pi\sim 1$.
\begin{figure}[h] \centerline{
\epsfxsize=12cm\epsfbox{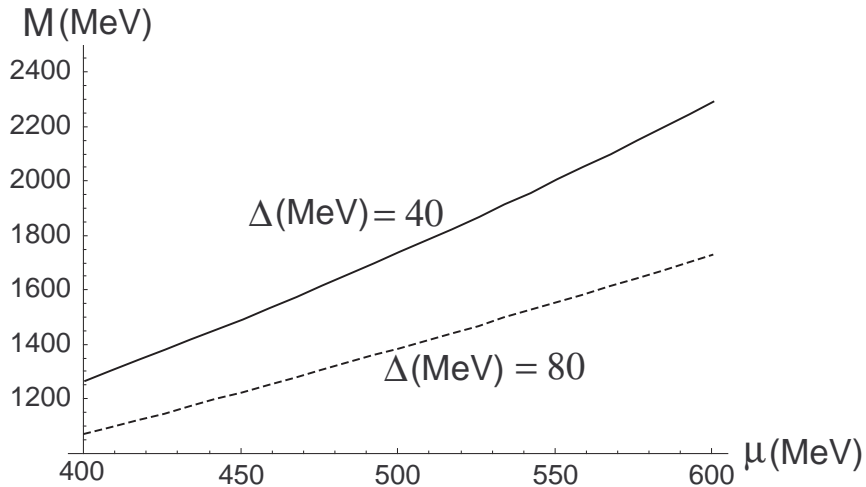} } \caption {{ \it The
soliton mass at finite density in the CFL phase as a function of
the baryonic chemical potential $\mu$, for two values of the gap
$\Delta$.\label{fig:1} }}
\end{figure}The soliton mass $M$ reported in Fig. \ref{fig:1}
corresponds to the classical solution, and does not take into
account $SU(3)_f$ breaking corrections or excitation energies such
as, for example, those corresponding to the pentaquark states.

 Fig. \ref{fig:1}
shows that around $400 $ MeV the soliton mass is about 1200 MeV,
which, in the light of the quark-hadron continuity, is in the right
ball-park, see the discussion in the next Section. It also shows
that at weak coupling, i.e. at larger values of the chemical
potential the  mass increases and the soliton effectively decouples.
Notice that the dependence of the soliton mass on the gap, at least
in this range of $\mu$ and $\Delta$,  is rather weak due the very
small coefficient in front of $\mu^2/\Delta^2$. In
\cite{Abud:1985wp} corrections due to higher derivative terms have
been discussed arguing that they should be small, however, we would
like to discuss here the validity of our approximation by looking at
the size of the soliton. Using the results obtained in
\cite{Adkins:1983ya} we find for the isoscalar mean radius (which
can be roughly assumed as the size of the instanton) \be r_0\approx
\frac{2.11}{2vF_Tg_s\sqrt{\lambda}}\ee On the other hand our
effective lagangian is valid up to energies lower than the gap
$\Delta$. Therefore, in order to describe correctly the soliton by
means of our effective lagrangian, one should have
$2vF_Tg_s\sqrt{\lambda}\ll 2.11\Delta$. We have studied this
condition by varying $\mu$. Since the mass of the soliton increases
with $\mu$ it decouples at high values of the chemical potential and
we do not expect to get a good description of the soliton in this
regime. Let us now consider smaller values of $\mu$ for which
$\lambda\approx 1$ (in practice this means $\mu\lesssim 10\Delta$).
We get the condition $1/r_0\approx 0.4\mu$. If we use
$\mu\approx400~MeV$ and $\Delta\approx 80~ MeV$, the result is
$1/r_0\approx 2\Delta$. Strictly speaking our description is not
valid up to this energy. However, in reference
\cite{Casalbuoni:2000na} we have shown that within the same
approximation one can evaluate the mass of the gluons (of order
2$\Delta$) in the CFL phase within a 30\% with respect to the exact
value. Therefore we can hope that  the same approximation holds at
the same level also in the present case. Clearly a better
approximation would be obtained by introducing higher derivatives in
the expansion. Let us estimate the error we are doing neglecting
them. To this end we will vary the function $F(r)$, taking into
account that also for the varied function, $F(0)=\pi$, which gives
the right topological number, and that, for $r>>r_0$, the new terms
are negligible. Therefore we have chosen to vary $F(r)$ in two ways.
In both cases we vary continuously $F(r)$ within the interval
$(0,2r_0)$ by keeping $F(0)$ and $F(r)$ fixed, for $r>2r_0$. In the
first case we increases the value of $F(r)$ at $r_0$ by 50\%,
whereas in the second case we take it 1/2 of the original value. The
results are the following: the mean radius increases of about 30\%
in the first case, whereas it is reduced by 50\% in the second one.
On the other hand in both cases the mass of the soliton increases of
about 30\%. This result follows from our estimate which is a lower
bound for the soliton mass since it is obtained by a variational
procedure. Therefore, our estimate is that the error we are
performing should not be higher than 30-50\% and that our results
should be qualitatively robust.

Again we remark  that, within our approximation, we have obtained
a very well defined expression for the soliton mass in the CFL
phase, containing no arbitrary parameters.
\section{Discussion and conclusions}
\begin{figure}[h] \centerline{
\epsfxsize=14cm\epsfbox{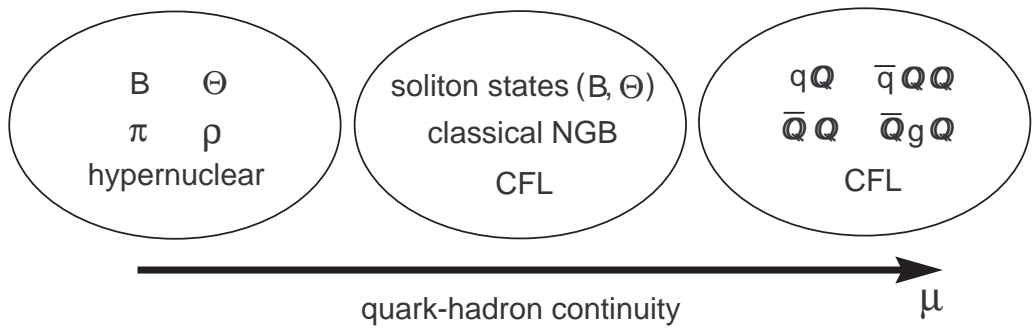} } \caption {{ \it A
cartoon depicting the transition from the CFL to the hypernuclear
phase. \label{fig:2} }}
\end{figure}

In  Section \ref{sec:1} we have shown that at energies close to
the Fermi energy $E_F$ and much smaller than $E_F+\Delta$ the
fermions decouple and the relevant degrees of freedom are the
Nambu-Goldstone bosons that can be thought of as
$\overline{\qq}\qq$ bound states. As discussed in the introduction
 we expect that in the CFL phase, besides these states, both quarks
and gluons become dressed, thus producing states such as $q\qq$,
$\overline{\qq} g \qq$, etc. The very high density CFL case is
depicted on the right-hand corner of the cartoon in Fig.
\ref{fig:2}.
In  Section \ref{sec:2} we have  shown that by taking the
next-to-leading order in the gluon decoupling process we get a
Skyrme term in the low energy lagrangian. Therefore the theory
predicts soliton states with the same quantum numbers of baryons.
However, being at weak coupling, the solitons have large masses
($M\approx 1/g_s$). As a consequence we expect the solitons to
decouple at the CFL densities. At lower densities  QCD coupling
gets stronger and the soliton mass  decreases. At these
intermediate densities we expect the low energy physics to be
still described by the chiral lagrangian, but with the soliton
states entering in to play. This correspond to the central part of
Fig. \ref{fig:2}. As discussed in the previous Section, this is
also the region where we expect that our approximation is valid.

 By decreasing the density one
should  go smoothly to the hypernuclear phase where the physical
states are pions, vector mesons and baryons (with the further
singlet state $H\approx udsuds$ corresponding to the Goldstone
boson $\phi$), as shown in the left corner of Fig. \ref{fig:2}.
Therefore the transition from CFL to the hypernuclear phase
appears completely smooth and without phase transitions. From this
point of view the existence of pentaquark states seems completely
natural. In fact, in the high-density limit, as we have seen,
quarks live in a dense medium made of diquark condensates.
Therefore a quark can bind a given number of diquarks. In
particular, one can form a bound state of the type $\bar q\qq\qq$,
that is a pentaquark. This same object is naturally described as a
soliton, and therefore it is expected to exist also in the
intermediate region and, by the quark-hadron continuity argument,
in the hypernuclear phase. Strictly speaking we cannot say that
this state persists also through the transition from the
hypernuclear to the nuclear phase, but this hypothesis appears to
be very natural. Some support to these qualitative ideas comes
from the numerical results of Section \ref{sec:2}. Using the low
energy parameters, as derived from the high-density limit, in the
intermediate density region, $\mu\approx 400\div 500~MeV$ we get
the right order of magnitude (see Fig. \ref{fig:1}), that is
$1.1\div 1.7~GeV$ varying the gap between 40 and 80 $MeV$. In the
same vein we can comment briefly about the expected width for the
lowest lying pentaquark state. In order for it to decay into a
baryon and a kaon a breaking of diquark condensates should be
produced. Of course this is very unlikely to happen at high
density.  One may expect that this feature survives going all the
way down to the nuclear phase.

\acknowledgements

We thank the Institute for Nuclear Theory at the University of
Washington for its hospitality during the completion of this
paper.


\end{document}